\documentclass[journal,10pt]{IEEEtran}

\ifCLASSINFOpdf
\else
\fi
\usepackage{graphicx}
\usepackage{epstopdf}
\usepackage{stfloats}
\usepackage{bm}
\usepackage[cmex10]{amsmath}
\usepackage{algorithmic}
\usepackage{algorithm}
\usepackage{array}

\usepackage{mdwmath}
\usepackage{mdwtab}
\usepackage{multirow}

\usepackage{amsfonts}

\usepackage{amssymb}
\usepackage{balance}

\usepackage{color}

\begin{document}
%
\title{Physical Layer Anonymous  Precoding: \\ The Path to  Privacy-Preserving Communications}

\author{Zhongxiang Wei,~\IEEEmembership{Member,~IEEE,}
            Christos Masouros,~\IEEEmembership{Senior Member,~IEEE,}
            H. Vincent Poor,~\IEEEmembership{Life Fellow,~IEEE,}
            Athina P. Petropulu,~\IEEEmembership{Fellow,~IEEE,}  
            and Lajos Hanzo,~\IEEEmembership{Fellow,~IEEE}

\thanks{Zhongxiang Wei is with the College of Electronic and Information Engineering, Tongji University, Shanghai, China. Email: z\_wei@tongji.edu.cn  }
\thanks{ Christos Masouros is with the Department of Electronic and Electrical Engineering at the University College London, London, U.K. Email:  c.masouros@ucl.ac.uk}
\thanks{H. Vincent Poor is with the Department of Electrical and Computer Engineering, Princeton University, Princeton, NJ, USA. Email: poor@princeton.edu}
\thanks{Athina P. Petropulu is with the Department of Electrical and Computer Engineering, Rutgers University, NJ, USA. Email:
athinap@rutgers.edu}
\thanks{L. Hanzo is with the School of Electronics and Computer Science, University of Southampton, Southampton, U.K. Email: lh@ecs.soton.ac.uk}

}

\maketitle

\begin{abstract}
Next-generation systems aim to increase both the speed and responsiveness of wireless communications, while supporting compelling applications such as  edge/cloud computing, remote-Health, vehicle-to-infrastructure communications, etc.
As these applications are expected to carry confidential personal data, ensuring user privacy becomes a critical issue.
In contrast to traditional security and privacy designs that aim to 
prevent confidential information from being eavesdropped upon by  adversaries, or learned by unauthorized parties, in this  paper we consider designs  that mask the users’ identities during communication, hence resulting in anonymous communications. 
In particular, we examine the recent interest
in physical layer (PHY) anonymous solutions. 
This  line of research departs from conventional higher layer anonymous authentication/encryption and routing protocols, and judiciously manipulates the signaling pattern of transmitted signals in order to mask the senders' PHY characteristics. We first discuss the concept of anonymity at the PHY, and  illustrate a strategy that is able to unmask the sender’s identity by analyzing his/her PHY information only, i.e., signalling patterns and the inherent
 fading characteristics.
Subsequently, we overview the emerging area of anonymous precoding to preserve the sender's anonymity, while ensuring high receiver-side signal-to-interference-plus-noise ratio (SINR) for  communication.
This family of anonymous precoding designs represents a new approach to providing anonymity at the PHY, introducing a new dimension for privacy-preserving techniques.

\end{abstract}

\IEEEpeerreviewmaketitle

\section{Introduction}


The development of 5G communications
has deeply influenced the era of  computing, storage and communications. 
The emerging applications are expected to carry confidential personal data, such as private data offloading in edge computing, real-time monitoring of vital physiological signals in remote-Health systems, e-Payment/Voting in smart city.
Hence, preserving the confidentiality of data becomes imperative in both commercial and military applications \cite{Bloch2021An},
where threats arise from two main aspects, namely security and privacy. 
Traditional research has focused  on secure design for averting  eavesdropping, where the aim is to prevent confidential information from being  exploited by external eavesdroppers (Eve)s. Hence,  there is extensive literature on cryptography, authentication, multiple-input and multiple-output (MIMO) beamforming, artificial noise, cooperative jamming, and radio frequency fingerprinting, spanning from the upper layers to the physical layer (PHY) of networks \cite{Dong2010Improving} \cite{Roy2020RFAL} \cite{Xiao2021Rein}. These extensive security-related 
solutions facilitate secure communications among legitimate parities, while ensuring that the signals cannot be captured by external Eves.

Naturally, cyber-threats also jeopardize privacy. While the data carried in emerging applications is becoming more personal, 
privacy leakage pervades all  applications
that require a user to release data in order to receive utility. 
In general, there are two  design principles in privacy protection, i.e, masking data itself or part of it, or concealing the users' identities for the intended receiver.
The former line of research seeks to control information leakage, and  strikes a compromise between privacy and utility, where  mutual information is used as a privacy measure  from the perspective of information theory \cite{Bloch2021An}.
Similarly, divergence-based metrics such as the  variation distance between the a \textit{priori} and a \textit{posteriori} distributions of the disclosed data are used as information leakage measures in the design of differential privacy \cite{Shkel2012Secrecy}.

A separate line of research focuses on guaranteeing the communication quality towards legitimate parties, while concealing the senders' identities during communication. This is known as anonymous communication \cite{Chen2017Lightweight}.
A typical example of  anonymous communication is found in cloud/edge computing, where the users wish to get their personal data processed at the cloud or edge. Nevertheless, it is expected that the receiver at the edge, i.e., the access point (AP), only processes the data without the capability of extracting the user's identity, such as his/her name, account, and other unique user-specific information.
Another example appears in e-Health applications, where patients wish to anonymously share their physiological information, but all the  private information, such as the patients' identities, must remain unknown at the receiver of the medical applications. Similar demand can be also found in anonymous e-Payment/Voting. 
In a nutshell,  anonymity-preserving techniques, where the receiver should only be able to process the data without the knowledge of the users' identities and participation, have become imperative in modern communications.

\begin{figure*}
	\centering
	\includegraphics[width=4.9 in]{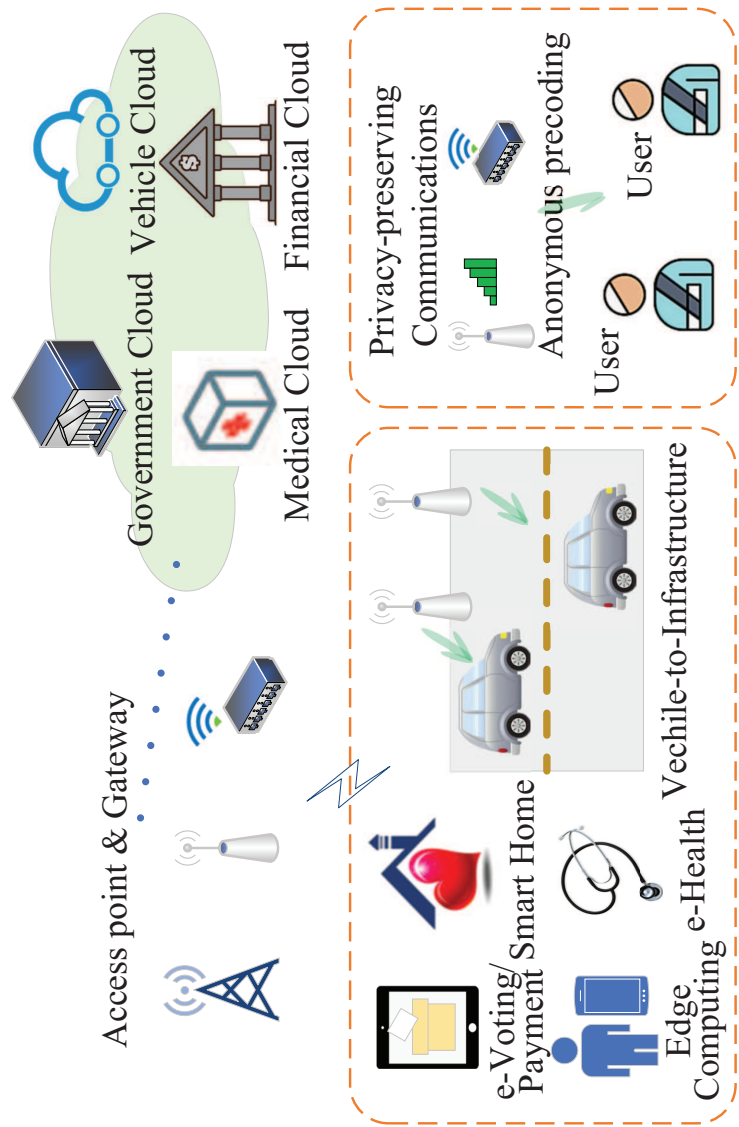}
   \caption{In emerging  applications, a high level of senders' anonymity is preferable through the upcoming paradigm of  privacy-preserving techniques at the PHY.  This new family of anonymous precoding techniques have been developed to mask senders' identities  against the receiver's  sender detection behavior, and meanwhile guarantee high receiver-side quality for communication.}
    \label{fig:ehealthscenario}
    \end{figure*}

At the upper layers of networks, a curious receiver may be able to extract the associated user ID during the authentication and encryption processes, or observe the user-specific characteristics of data traffic along the routing path to infer the real identity of the sender. Accordingly, the existing anonymity-preserving techniques that reside at upper layers can be classified into anonymous authentication, anonymous encryption and anonymous routing protocols, which are briefly summarized in Table I.
Higher layer anonymous authentication
and encryption schemes have been proposed for cellular networks,  
Radio Frequency IDentification \cite{Chen2017Lightweight}, wireless body area
networks \cite{Liu2014Certificateless}, vehicle-to-infrastructure communications  and for prototype design \cite{Emura2016Secure}. 
The design philosophy of upper layer anonymous authentication/encryption is to apply group/ring signatures or anonymous account indices \cite{Liu2014Certificateless}  for authentication, as well as for constructing communication links and encryption. Since the users only share their so-called pseudo accounts with the service provider, their real IDs are not exposed. In this way, a potential adversary remains unaware of the users’ real identities and hence fails to identify them.
On the other hand, the fundamental principle of  anonymous routing  is to conceal the user as well as the related routing paths  via a number of encrypted layers, relying on an onion-structure;
this structure has found wide applications in the Internet and ad-hoc networks \cite{Sakai2019On}. 
It is striking that, while numerous PHY solutions exist for security, there is a dearth of PHY solutions for anonymity. 

Higher layer anonymizing techniques have  some drawbacks restricting their practical implementations, i.e.,

\begin{itemize}
    \item  Since the existing anonymous authentication/encryption schemes  generally  rely on public-key encryption, asymmetric encryption, identity-based encryption, fully homomorphic encryption, cryptographic primitive, etc., they  require additional key distribution, agreement and maintenance processes. This may be restrictive in many emerging scenarios  due to their excessive complexity and latency. Although  small keys  have been proposed based on elliptic curve cryptosystems, the communicating parties still need  to verify the certificates of others, and a pool of certificates is required by the certification authority for storing the legitimate parties’ keys. 

    \item The existing anonymous routing protocols require  help from external cooperative agents/routers.  The prolonged routing path and the data re-directing mechanism  inevitably increase latency. Also, since the anonymous routing protocols give trust to the cooperators along the routing path, they may be  vulnerable to internal attacks by curious cooperating members. Further, the cooperators need to stay online during the underlying process thus  consuming power.
    
    \item As the existing anonymous authentication, encryption and routing designs are  designated for the upper layers, they assume that a privacy-preserving link has been established at the PHY.  In fact, the hitherto neglected PHY  contains critical information that can be analyzed for extracting the senders' identities. For example, when an anonymously authenticated/encrypted user  sends a signal via its wireless channel, the receiver can  analyze the signalling patterns of the received signal.
    Based on the channel fading characteristics, the receiver may be able to determine the origin of the signal. 

\end{itemize}


\begin{table*}
\centering
	   \caption{A Brief Summary of The Existing Research, from the Perspectives of Security and Anonymity.}
	\includegraphics[width=6.2 in]{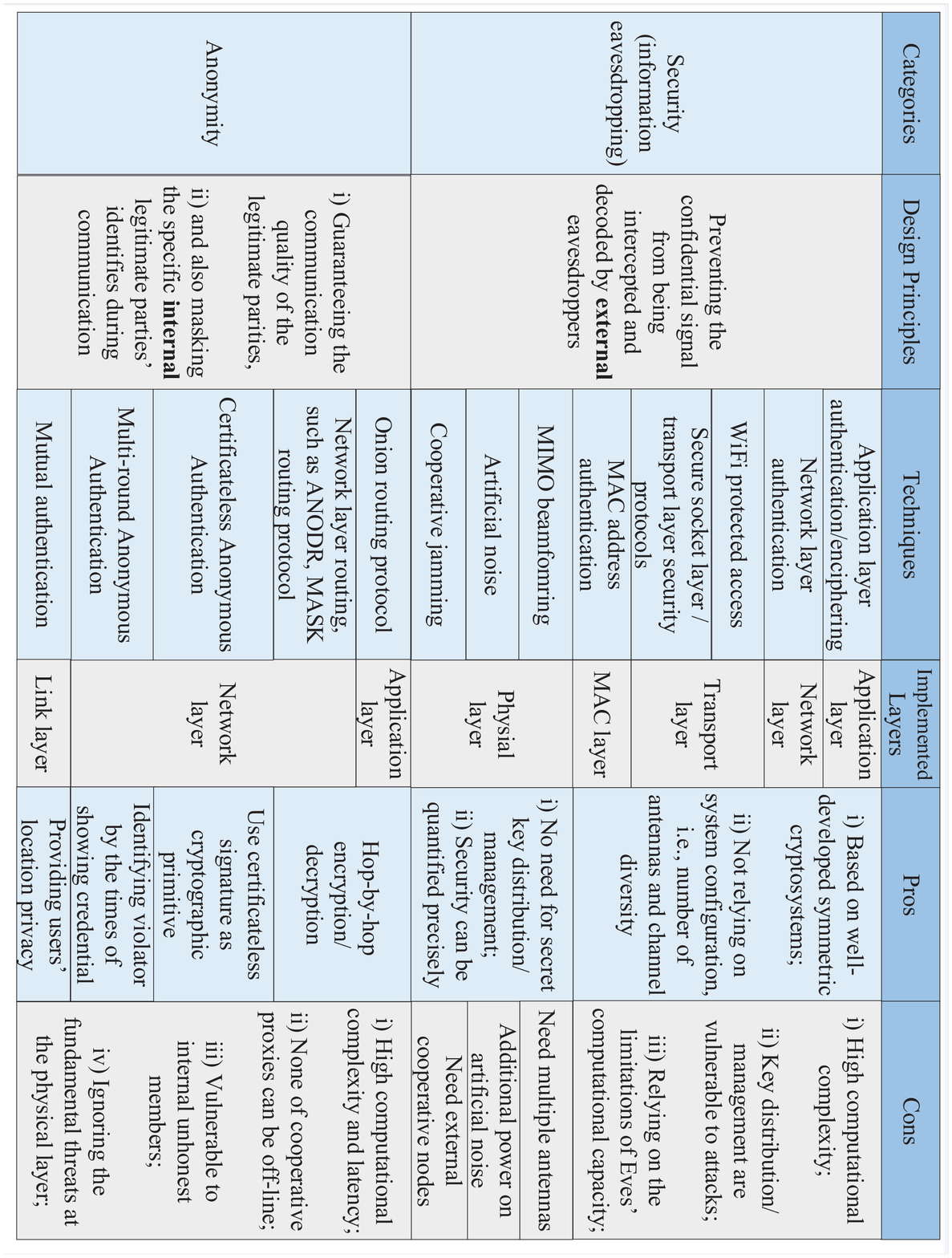}
    \label{fig:summary} 
\end{table*}

Lack of defense at the PHY may compromise the sender's privacy.
Hence,  using complementary anonymity-preserving techniques that reside at the PHY is important.
However, attaining PHY anonymity is particularly challenging, as it entails the conflicting aims of, on one hand, ensuring high signal quality for data detection at the receiver, and on the other hand, ensuring that the identity of the sender remains hidden from the same receiver.
The objective of this article is to introduce the emerging field of  PHY anonymity. Commencing from the fundamentals of PHY anonymity, we discuss the issues of  sender detection and how it can  be applied at the receiver to break the sender's anonymity. Subsequently, the corresponding anonymous precoding to be used at the sender side is reviewed, followed by a range of open challenges and our conclusions.

\section{Performance Metric and Fundamentals of Anonymity}

Anonymity can be quantified by an entropy-based metric  \cite{Chou2007An}. 
Assuming that there are $K$ possible senders whose channels towards the receiver are known, the receiver estimates for each potential user $k$ a probability  of being the real sender. 
The highest level of anonymity is achieved when all  possible senders are  equally probable,   resulting in the maximal anonymity entropy.
To break the sender's anonymity,  the  receiver should correctly declare the real sender with a high probability, while computing low probabilities for all other potential senders.
On the other hand, at the sender side,  an appealingly simple anonymity-preserving technique is to manipulate the transmitted waveform, so that the sender-detection error rate (DER)  at the receiver deteriorates, which will be detailed in Section IV.
Naturally, anonymity should be achieved without any loss of the reception quality.

Since the users wish to communicate with the receiver in an anonymous manner, they could transmit data without notifying the receiver of their identities, such as their MAC/IP addresses. Hence,  the receiver  would have knowledge of the  received  signal only.
Before the receiver would attempt to extract the origin of the received signal, it would have to first sense the presence of an incoming signal. 
The detection of the signal's presence would rely on ubiquitous  signal-to-noise ratio (SNR), eigenvalue or feature based energy detector, which has been widely investigated in cognitive radios to sense the availability of spectrum. 
Once an incoming signal is sensed, the receiver would apply its detection scheme to identify the sender. 

\section{Sender Detection Strategy }

\begin{figure}
	\centering
	\includegraphics[width=3.4 in]{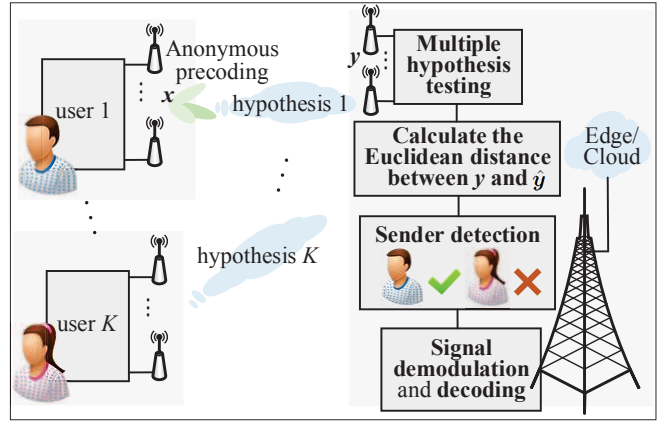}
    \caption{ A generic receiver-side signal processing thread by the curious receiver, where sender detection is performed to unmask the real sender using a multiple hypothesis test.}
    \label{fig:anonymous0828} 
\end{figure}

Let us consider a typical uplink multiuser MIMO, where the receiver and users are equipped with multiple antennas.
Regardless of the precoders applied at the sender, the transmitted signal  propagates through the wireless channel, which is a unique, user-specific PHY identity of the specific sender, before it  reaches the receiver.
Hence, under the time-division-multiple-access  premise, the detection of the sender's identity is equivalent to the identification of the propagation channel of the received signal.
Since the recipient has sensed the amplitude and phase of the received signals originating from a specific member of the set $\mathbb{K}$,   sender detection  can be formulated as a multiple hypothesis testing (MHT) problem. In particular, each possible sender is associated with a specific hypothesis.



Considering  non-line-of-sight transmission experiencing flat-fading on each carrier, since the receiver has  knowledge of channel state information (CSI) of the possible senders, the receiver can estimate the transmitted vector $\bm{x}$ (which is also the precoded signal at the sender) by multiplying the received signal with the pesudo-inverse of a hypothesized channel. 
Then, the estimated version of the transmitted signal is left-multiplied by the hypothesized channel  to imitate that it propagates through the hypothesized MIMO channel, resulting in a  re-constructed signal $\hat{\bm{y}}$.
If the received signal indeed had propagated through the  channel considered, the re-constructed signal based on the hypothesized channel would have the smallest Euclidean distance from the real received signal. 
Motivated by the above observations, the receiver's detection strategy is outlined as follows. The receiver first estimates the transmitted vector and constructs the hypotheses  $\hat{\bm{y}}$ relying on all the possible candidates' channels. Finally, the hypothesis having the smallest Euclidean distance to the actual signal is declared as the real sender \cite{Wei2020Fundamentals}.
This Euclidean distance based detection relies on the disparity of the small-scale channel fading among the potential senders. 
However, there are scenarios, for example in high-frequency (e.g., millimeter wave) regimes, where even if some users are closely spaced, their small-scale fading are different, and thus  the receiver can unveil the sender with high probability.

The above approach shows that, using  the signal propagation model and  knowledge of the channel, the receiver can determine who the sender is, in a fashion that  is completely independent from the decryption at higher layers. 
In other words, although the transmitted data can be anonymously authenticated/encrypted at  higher layers by the sender, the  receiver is  able to directly unveil the sender at the PHY  \cite{Wei2020Fundamentals}.
To this end, the classical precoders are no longer anonymous in nature, allowing a leakage of the sender's identify at the PHY.
For example, by the well-known singular value decomposition (SVD), signal-to-interference-plus-noise ratio (SINR) balancing, and transmission power minimization precoders, 
the additional step of channel equalization 
has to be performed at the receiver  for de-modulation. Since a  receiver-side equalizer exploits the  knowledge of the real sender's  channel, 
this knowledge can be leveraged by the receiver to extract the sender's identity.

\section{Anonymous Precoding Design}

Now,  we examine the family of anonymous precoding techniques that are applied at the sender to hinder the receiver's  detector.

The aim of an anonymous precoder is to conceal the sender's identity, and at the same time provide high reception quality \cite{Wei2020Physical}. 
Note that the anonymous precoding technique may be viewed as an outer layer of the existing group/ring signature and anonymous account index based privacy preservation techniques. Explicitly, the existing techniques allocate users pseudo-IDs for authentication, as well as for constructing links and encryption, while anonymous precoding directly improves privacy preservation during communication phase.
Interestingly,  implementing  sender anonymity conflicts with the design of the receiver-side equalizer, which can be proven via the following counter example.
If the communication quality can be enhanced by the receiver-side equalizer, no anonymity can be achieved, as the equalizer exploits the real sender's  channel. On the other hand, if anonymity is guaranteed and the sender's identity is masked, the receiver is unable to determine the correct propagation channel, indicating that receiver-side equalization is impossible.
Hence, the receiver will have to directly demodulate and decode the per-antenna multiplexed data without a reception equalizer; here  the MIMO channel between the sender and receiver acts as a multiuser multiple-input and single-output (MISO) channel, as we have to treat each antenna of the receiver  individually  and consider per-antenna  SINR constraint for multiplexing  data streams.
In the following, we review a pair of anonymous precoder designs \cite{Wei2020Fundamentals} \cite{Wei2020Physical}.

\subsection{Interference Mitigation Based Anonymous Precoder}

The philosophy of the interference mitigation (IM)-based anonymous precoder is that of  manipulating the transmitted waveform for masking the user-specific information (e.g., CSI) during data transmission, and at the same time
strictly suppressing the inter-antenna interference (IAI) using the techniques of  suppressing the multi-user interference in MISO systems \cite{Wei2020Fundamentals} for communication. 
All  active users  send pilot signals to the AP and channel estimation is performed at the AP during   training phase. Then the CSI is fed back to the users to be used for  precoder design, as in generic MIMO communications. Although the AP can know the users' CSI in its cell, it does not  jeopardize the anonymity aims,
as the anonymous precoder still succeeds in preventing the AP/server from linking the data received to  correct user ID and CSI during communication phase.

Indeed there has been extensive research  on anonymity-agnostic MISO precoder design, relying on total power minimization under minimum SINR requirement,  SINR balancing under power budget constraints, maximization of the total energy-efficiency under per-user SINR constraint, etc \cite{Li2018Interference}. These problem formulations are generally given as convex or non-convex second order cone programming (SOCP) exercise, where the resultant problem can be mathematically transformed and handled by semi-definite programming (SDP).
When considering the senders' anonymity, 
an additional anonymity constraint should be imposed on the precoder.
Again, the receiver first re-constructs the different hypotheses  and considers the  one having the smallest Euclidean distance to the actually received signal  as the real sender. Hence, to preserve anonymity, the sender can randomly select another candidate from the set $\mathbb{K}$ as an alias sender. That is, the anonymous precoder needs to be appropriately designed to guarantee that the re-constructed signal based on the alias's channel  is equivalent to the signal corresponding to the real sender's channel. As a result, the real sender and the alias become equally likely senders from the receiver's perspective, and hence the receiver fails to declare the correct one.
The anonymity constraint imposed is linear in nature, which can be simply incorporated in a conventional MISO precoder without breaking the convexity of the problem formulation \cite{Wei2020Physical}. 
Hence, the anonymous precoder can be readily obtained, which guarantees a high reception quality while preserving the anonymity of the sender.

\subsection{Interference Exploitation Based Anonymous Precoding}
\begin{figure*}
	\centering
	\includegraphics[width=6.3 in]{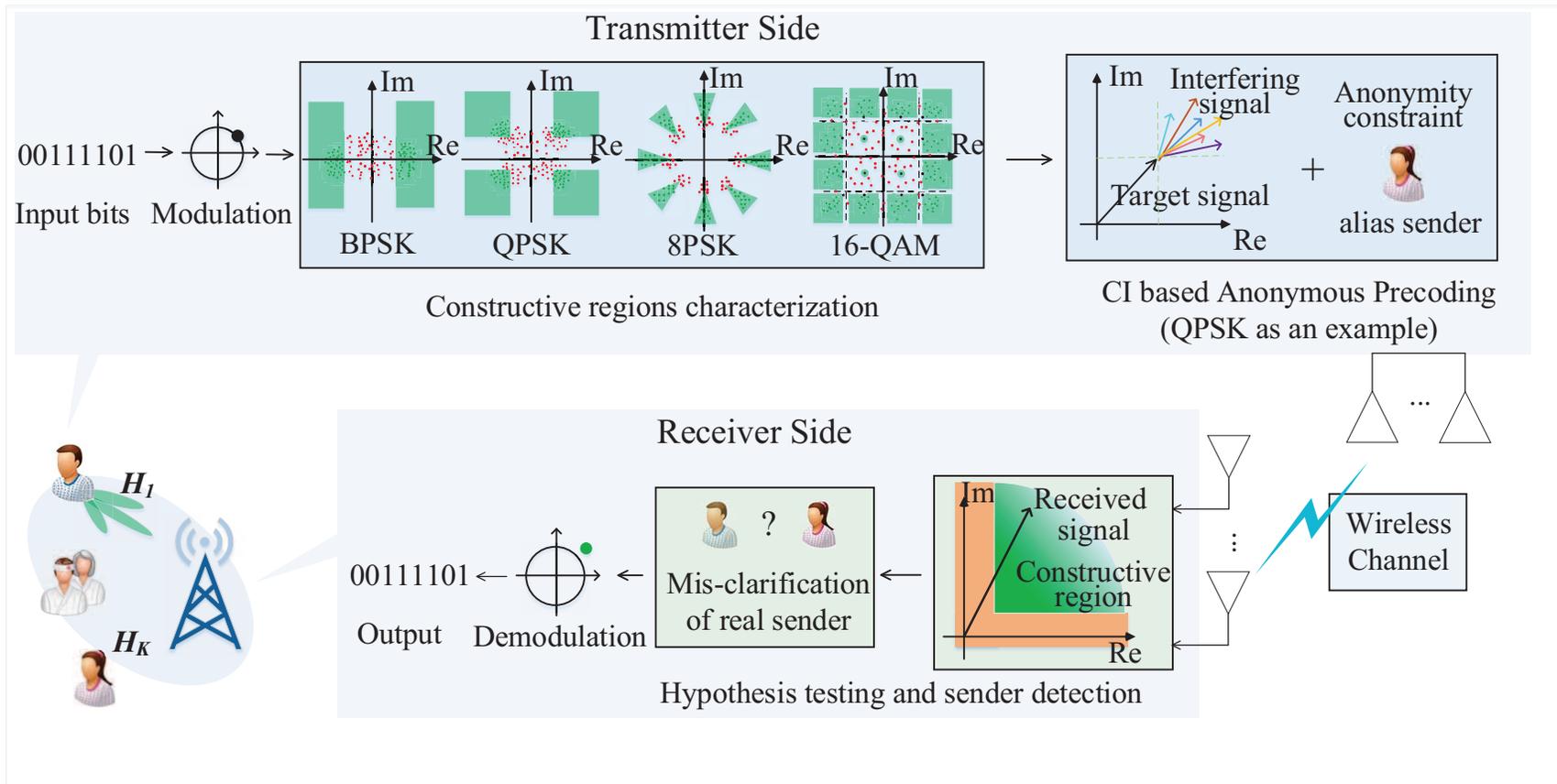}
    \caption{A generic CI anonymous precoder guideline. At the sender,  CI regions are first exploited, and then the symbols are precoded under the CI and anonymity constraints. 
    At the receiver, the users are tested as equally suspicious senders, and hence the sender detection is disabled. Furthermore, since the  per-antenna received signal falls into CI regions (green area),
    conventional  demodulation and decoding can be directly performed to obtain the output.}
    \label{fig:anonymous0709} 
\end{figure*}

In the above section, we have discussed the fundamentals of the IM-based  anonymous precoder, where the inter-antenna interference is considered as a harmful element that  degrades the system performance.
As a result, the IM-based precoder has to restrict the received symbols within a certain region around the nominal point in the  signal
constellation. In fact, the idea of suppressing inter-antenna interference is only optimal from a statistical perspective, where the input  is considered as an infinite-cardinality Gaussian signal and only the spatial correlation of the wireless channels is exploited. 
However, since the  symbols to be transmitted are known by the sender, we can  jointly exploit the spatial correlation among the channels and the intended symbols, based on the concept of constructive interference (CI) \cite{Christos2013Known}.
Hence, the inter-antenna interference has the potential to be exploited as a useful element to push the per-antenna received symbol  away from the detection thresholds of the signal constellation. The resultant increased distance with respect to (w.r.t) the detection threshold directly  benefits the per-antenna reception quality. 
Let us first demonstrate the concept of CI, and  then exploit it for our anonymous precoder design. 


\subsubsection{Constructive Interference Based Precoding}
We illustrate a generic CI-based anonymous precoder design
in Fig. 3  for both binary phase shift keying (BPSK), as well as  for quadrature phase shift keying (QPSK), and quadrature amplitude modulation (QAM), etc. 
First, given the  constellation adopted, interference characterization can be performed.
The CI regions of each constellation are
denoted by the green areas of Fig. 3. To be specific, given the decision boundaries of each modulation
constellation, the interference is said to be constructive if it
pushes the desired signals away from the decision boundaries
while it is said to be destructive if it pulls the  signal
close to or even across the decision boundaries.  
For example,   the decision thresholds are the real
and imaginary axes for  QPSK, and hence the interference becomes beneficial if it pushes the desired signals away from both the real and imaginary axes. The characterization of the CI regions can be  defined for higher order modulations, such as 8PSK and  16QAM \cite{Wei2020Multi}, in a similar vain.
Hence, we are able to  make the IAI always beneficial, where the  per-antenna received signal  is directly shifted within the CI regions. By utilizing the inter-antenna interference as a beneficial element, the reception quality can be further improved over that of the IM-based precoder.

\subsubsection{CI Based Precoding with Anonymity Constraint}

A key advantage of the CI-based precoding w.r.t anonymity is that, since the received signal is directly shifted into the CI regions, the receiver can simply demodulate the signal, based on its amplitude and phase.
As a result, the need for channel equalization at the receiver-side is removed, and there is no need to identify the user's channel that would reveal the sender identity.
Hence, the foundation of the CI-based precoding is particularly well suited for anonymous transmission, which is also able to exploit the IAI interference as a helpful component without loss of anonymity.

For illustration, Fig. 3 portrays a generic CI-based anonymous precoder. For employing CI  
preserving the sender's anonymity, it is essential to impose an additional anonymous constraint to manipulate the pattern of the received signal to mask the sender's PHY characteristics, as we discussed for the IM-based  anonymous precoder. 
In fact, it has been exhibited that the CI precoder itself is a linear precoder, and the CI-related optimization is generally convex in nature \cite{Wei2020Multi}. Hence,  imposing an additional convex anonymity constraint does not break the convexity  and the feasibility of the  optimization. 


\begin{figure*}
	\centering
	\includegraphics[width=6.3 in]{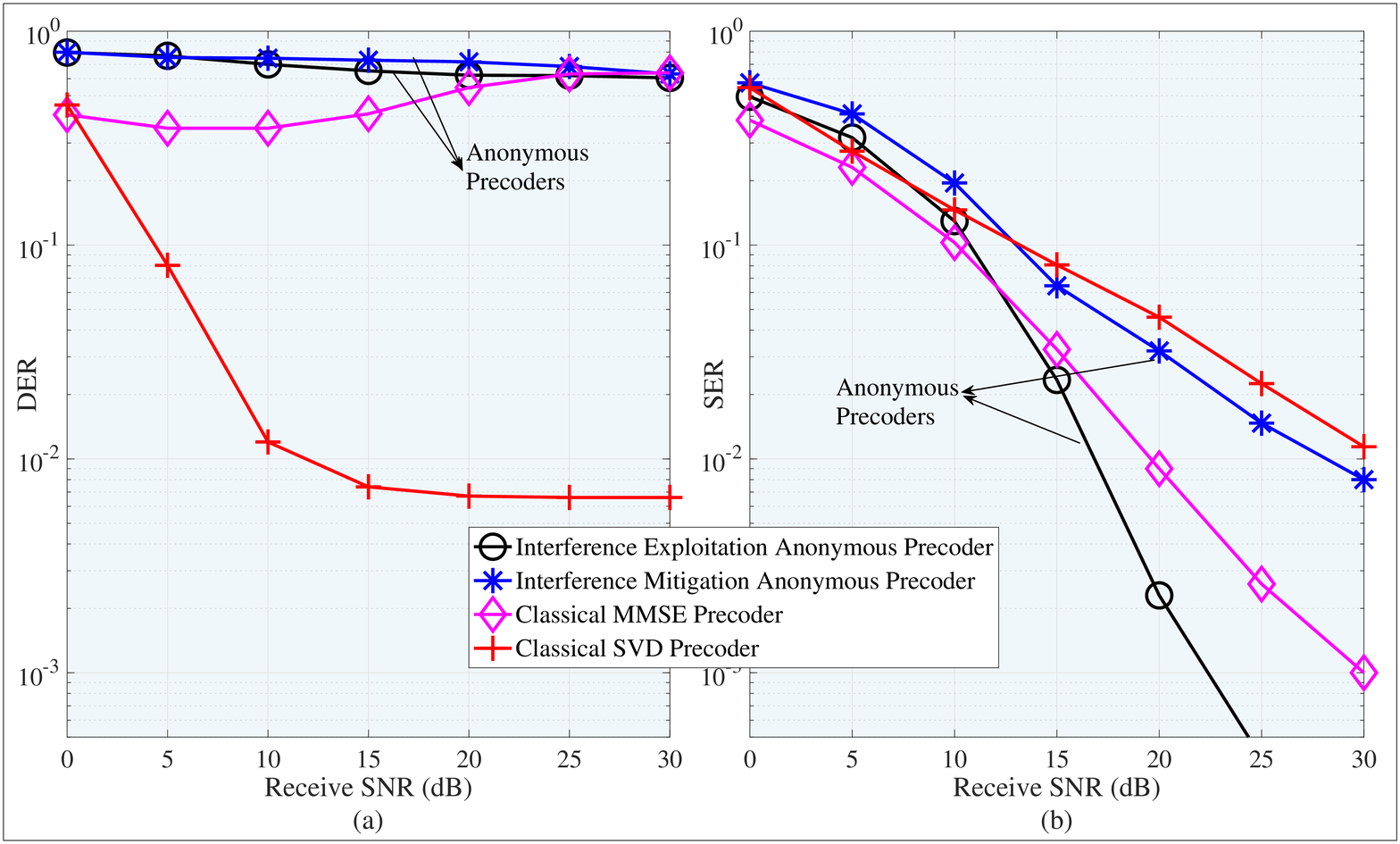}
    \caption{
     (a) Anonymous precoders scramble the receiver’s detection for preserving the sender’s anonymity, leading to a high DER at the receiver. (b) Low SER is also guaranteed by the anonymous precoders compared to the benchmarks \cite{Wei2020Physical}.}
    \label{fig:anonymous_mag} 
\end{figure*}

\subsection{Comparative Evaluation}

In Fig. 4, the DER and symbol error rate (SER) vs. the SNR trends are demonstrated, benchmarked by the MMSE and SVD precoders \cite{Li2018Interference}. We assume that the hypothesis set $\mathbb{K}$ includes 5 candidate users, where the sender in each time slot (block) is randomly generated. The senders and receiver are equipped with 10 antennas. The power budget is 1 Watt and QPSK modulation is used. It is observed that both the IM and IE anonymous precoders can provide anonymity for the sender, where the receiver’s DER is as high as 0.8 at high receive SNRs. By contrast, the SVD precoder shows the worst anonymity performance, as the receiver is able to identify the correct sender with a 0.01 DER at 10 dB SNR. For the MMSE precoder, the receiver’s DER exhibits a gentle trough. This is because at low receive SNRs, its detection performance is impaired by strong noise, but at high receive SNRs the operation of the MMSE precoder is reminiscent of that of the ZF precoder. Because no CSI knowledge is relied upon, anonymity is incidentally preserved. 
In Fig. 4(b), since the IE-based anonymous precoder is capable of exploiting the IAI as a desired component, the high DoFs at the sender results in an excellent SER performance, achieving 14 dB SNR gain over the standard MMSE and SVD precoders. Furthermore, the IM-based anonymous precoder can be generally designed by SDP optimization, which is known to have good performance at moderate/high SNRs. Hence, the IM-based anonymous precoder exhibits an SER close to that of the SVD precoder within the 0-12 dB SNR range, while in the region above 12 dB SNR, it outperforms the SVD precoder by about 3 dB. Thus, the anonymous precoders strike an attractive trade-off by providing high communication performance, while preserving the sender’s anonymity.

Note that since additional anonymity constraints are required for the design of IM-based and IE-based anonymous precoders, the improved anonymity is achieved at the cost of reduced DoFs in precoder design.
More importantly, the design principle of the anonymous precoding technique is to manipulate the transmitted waveform for masking the sender’s identity-related information. Since it does not require help from external proxies, nor does it rely on complex networking or dedicated data re-routing protocols, anonymous precoding techniques are readily applicable to small-scale networks that have simple architectures and protocols.



\section{Open Challenges and Future Research}

\textbf{PHY Anonymity Design with Practical Issues}:
Practical  implementations always suffer hardware impairments, and  acquiring accurate CSI may be difficult.
All these imperfections would jeopardise the grade of anonymity, which are particularly  pronounced  in   wearable sensors of e-Health applications, or portable devices of edge computing, that are built with low-cost devices.
Some pioneering impairment-aware secure PHY solutions do exist, but they cannot be directly applied for anonymous systems. 
The validation of anonymous precoding considering the above practical issues needs further research.


\textbf{The Interplay between Anonymity and Data Security in Heterogeneous Networking}:
The massive number of communication devices results in heterogeneous networks, including but not limited to WBAN, sensor networks, edge/cloud computing, etc. As mentioned in the Introduction, anonymity and data security have distinct objectives and design principles, while in practices, different users and applications may have heterogeneous anonymity and data security specifications. This requires the design of dedicated algorithms for studying the underlying trade-off between anonymity and data security.

\textbf{PHY Anonymity with Ultra Reliability and Low Latency  Applications}: 
Some emerging applications, such as augmented reality and haptic communications, require ultra-reliability and low-latency (i.e., URLLC), where short packet transmission has been considered as a key technique of meeting the stringent requirements of URLLC applications. Evidently, the joint optimization of anonymity, reliability, and latency is non-trivial, which may incur high complexity and overheads. How to preserve PHY anonymity without loss of reliability and increasing latency at acceptable degrees of complexity and overhead, remains an open challenge.

\textbf{PHY Anonymity from the Perspective of Intelligent Receiver}: 
Providing that abundant computing capability is attainable, the receiver is able to detect the signal sender with the aid of machine learning (ML), which may be hence regarded to as an intelligent-receiver.
Leveraging the concept of ML, an intelligent-receiver can analyze the
statistics of the signal and  develop an ML-enabled detection mechanism, thus improving its detection performance.
Thus, it is necessary to reconsider the anonymizing designs  of intelligent-receivers.

\section{CONCLUSIONS}
This article has discussed the privacy-preserving techniques at the PHY, aiming to provide a comprehensive overview for the concept of PHY anonymity.
Starting from the fundamentals of PHY anonymity,
we first have discussed the methodology of the sender detection  at the receiver, which enables the receiver to disclose the sender by only analyzing PHY information.  
Subsequently, we have examined the corresponding  anonymous precoders to provide the sender's anonymity while guaranteeing  a high receive-quality for communication.
The reviewed anonymous precoders judiciously manipulate the signalling pattern to  scramble the receiver's malicious detection behavior without harming the convexity and feasibility of precoding design, offering a
broad field of preserving anonymity for the
privacy-critical applications. 
Finally, a number of challenges on PHY
anonymity are envisaged.

\section*{Acknowledgement}
Z. Wei would like to acknowledge the financial support of the National Natural Science Foundation of China under Grant 62101384.
C. Masouros would like to acknowledge the financial support of the Engineering and Physical Sciences Research Council under
Grant EP/R007934/1.
H. V. Poor would like to acknowledge the financial support of the U.S. National Science Foundation under Grant CCF-1908308.
A. P. Petropulu would like to acknowledge the financial support of the ARO under Grant W911NF2110071.
L. Hanzo would like to acknowledge the financial support of the Engineering and Physical Sciences Research Council under Grants EP/P034284/1 and EP/P003990/1 (COALESCE) as well as of the European Research Council's Advanced Fellow Grant QuantCom under Grant 789028.

\begin{IEEEbiographynophoto}{Zhongxiang  Wei}
received his Ph.D. degree in electrical and electronics engineering from the University of Liverpool, Liverpool, U.K., in 2017. From March 2016 to March 2017, he was with the Institute for Infocomm Research, Agency for Science, Technology and Research, Singapore, as a Research Assistant. From March 2018 to March 2021, he was with the department of electrical and electronics engineering, University College London, as a research associate. In April 2021, he joined Tongji University as an associate professor. His  interests include PHY anonymity and MIMO systems design. He has served as a symposium/special issue co-chair/TPC member of IEEE
ICC, GLOBECOM, and ICASSP.
\end{IEEEbiographynophoto}

\begin{IEEEbiographynophoto}{Christos Masouros} is the Professor of Electrical and Electronic Engineering at UCL. He received his PhD in Electrical and Electronic Engineering from the University of Manchester, UK (2009). He has held a Royal Academy of Engineering Research Fellowship (2011-2016).
His  interests include wireless communications and signal processing with speciality on Large Scale Antenna Systems and Interference Exploitation. He is an Editor for IEEE TWC/TCOM, and has been an Associate/Guest Editor for IEEE COML/JSTSP.
He was the recipient of the Best Paper Awards in the IEEE GLOBECOM'15 and WCNC'19. 

\end{IEEEbiographynophoto}

\begin{IEEEbiographynophoto}{H. Vincent Poor} is the Michael Henry Strater University Professor at Princeton University, where he has been on the faculty since 1990. Prior to joining Princeton, he was on the faculty of the University of Illinois at Urbana-Champaign. He has also held visiting positions at several other universities, including most recently at Berkeley and Cambridge. His research interests are in the areas of information theory, machine learning and network science, and their applications in wireless networks, energy systems and related fields. Among his publications in these areas is the forthcoming book Machine Learning and Wireless Communications, to be published by Cambridge University Press. Dr. Poor is a member of the National Academy of Engineering and the National Academy of Sciences and is a foreign member of the Chinese Academy of Sciences, the Royal Society, and other national and international academies. He received the IEEE Alexander Graham Bell Medal in 2017.
\end{IEEEbiographynophoto}

\begin{IEEEbiographynophoto}{Athina P. Petropulu} is a Distinguished Professor of Electrical and Computer Engineering at Rutgers University. Her interests include radar signal processing and PHY security. She received the Presidential Faculty Fellow Award (1995) from NSF and the US White House, and the 2012 IEEE Signal Processing Society (SPS) Meritorious Service Award. She is IEEE and AAAS Fellow. She is co-author of the 2005 IEEE Signal Processing Magazine Best Paper Award, the 2020 IEEE SPS Young Author Best Paper Award and the 2021 Aerospace and Electronic Systems Society Barry Carlton Best Paper Award. She is currently President-Elect of the IEEE SPS.
\end{IEEEbiographynophoto}

\begin{IEEEbiographynophoto}{Lajos Hanzo}  (http://www-mobile.ecs.soton.ac.uk, https://en.wikipedia.org/wiki/Lajos\_Hanzo) (FIEEE'04) received his Master degree and Doctorate in 1976 and 1983, respectively from the Technical University (TU) of Budapest. He was also awarded the Doctor of Sciences (DSc) degree by the University of Southampton (2004) and Honorary Doctorates by the TU of Budapest (2009) and by the University of Edinburgh (2015).  He is a Foreign Member of the Hungarian Academy of Sciences and a former Editor-in-Chief of the IEEE Press.  He has served several terms as Governor of both IEEE ComSoc and of VTS.  He is also a Fellow of the Royal Academy of Engineering (FREng), of the IET and of EURASIP.
\end{IEEEbiographynophoto}

\ifCLASSOPTIONcaptionsoff
  \newpage
\fi

\end{document}